\documentclass[aip,reprint,amsmath,amssymb]{revtex4-1}

\usepackage{graphicx}% Include figure files
\usepackage{dcolumn}% Align table columns on decimal point
\usepackage{bm}% bold math
\usepackage[utf8]{inputenc}
\usepackage[T1]{fontenc}
\usepackage{enumitem}
\usepackage{tgschola}
\usepackage{hyperref,physics,caption,subcaption,xcolor}
\usepackage{etoolbox}
\usepackage{comment}
\usepackage{float,placeins}
\usepackage{tikz}

\makeatletter
\def\@email#1#2{%
 \endgroup
 \patchcmd{\titleblock@produce}
  {\frontmatter@RRAPformat}
  {\frontmatter@RRAPformat{\produce@RRAP{*#1\href{mailto:#2}{#2}}}\frontmatter@RRAPformat}
  {}{}
}%
\makeatother
\begin{document}

\preprint{AIP/123-QED}

\title{Nonlinear Frequency Shifts due to Phase Coherent Interactions in Incompressible Hall MHD Turbulence
}

\author{E.C. Hansen}%
\affiliation{Institute of Fusion Studies, University of Texas at Austin}
 \email{ehansen99@utexas.edu}
 
\author{Prerana Sharma}
\affiliation{Institute of Fusion Studies, University of Texas at Austin}
\affiliation{Physics Department, Ujjain Engineering College}

\author{S.M. Mahajan}
\affiliation{Institute of Fusion Studies, University of Texas at Austin}

\date{\today}
\begin{abstract}
Turbulence in the magnetized plasma is well understood to be the consequence of wave interactions. 
When the Hall effect is added to the minimum magnetohydrodynamics (MHD), the MHD waves become dispersive and different nonlinear interactions are expected.
The emergent turbulent state will thus be expected to be different.
For incompressible Hall MHD we develop a reduced model for wave-wave interactions concentrating on those processes that will lead to phase coherent modifications to the linear dispersion of a given wave. 
We show that these special interactions provide an amplitude-dependent contribution to the linear dispersion relation, which yields nonlinear frequency shifts. 
The resonance-driven frequency shifts are dominant and add damping or growth to the linear dispersion.
The damping/growth rates represent the nonlinear time scales for energy redistribution and can be used in conjunction with a conjecture like the ``critical balance’’ to estimate the energy spectral content.
\end{abstract}

\maketitle

\setlength{\parindent}{0pt}
\setlength{\parskip}{1em plus 0.5em minus 0.5em}
\section{Introduction}
The ability to support waves is one of the primary features that distinguishes magneto plasmas from neutral fluids; it readily translates into different nature of turbulence manifested in the respective models- magnetohydrodynamics (MHD) and hydrodynamics- used to describe them. 
In MHD, the Alfvén wave connects the oscillating fluctuations in the velocity and magnetic field perturbations.\cite{Alfven1942} 
Alfv\'en waves are a generic name for a host of relatively low frequency (lower than ion cyclotron frequency) wave motions accessible to a magneto-plasma. 
Due to their essential role in dictating plasma properties, Alfv\'en waves constitute a highly investigated research area. 
It is not our intention to survey this vast field but we do give here a list of representative references that the reader may consult. 
Alfvén waves play a fundamental role in determining the stability of and turbulence in magneto plasmas; they are also an efficient way of heating them.\cite{Ross1982,White2002}
In the solar wind, durable observed correlations between the magnetic and velocity field perturbations reflect Alfvénic activity.\cite{BelcherDavis1971,Klein2012}
A discrete spectrum has been found for Alfvén waves on the inclusion of the parallel electric field in a cylindrical plasma.\cite{Mahajan1984}
The particular manifestations of Alfvén waves in toroidal geometry have also received much attention, \cite{Huanchun1993,MettMahajan1992}
and of particular importance has been the destabilization of Alfvén eienmodes by energetic particles.\cite{Li1987,ChenZonca1995,Vlad1999,Heidbrink2008}
Many observations of Alfvén eigenmodes in tokamak plasmas and their impact on fast ion confinement and achieving fusion gain have been documented,\cite{Wong1999} 
and their appearance in optimized stellarators remains an active area of research.\cite{Kolesnichenko2001,Paul2025}

Most of the aforementioned studies deal with the linear wave dynamics of a fluid plasma system. However, the eventual impact of these waves on the plasma dynamics must be understood in a nonlinear, possibly, turbulent state. 
The latter, of course, originates in the interactions of these waves.
\cite{Iroshnikov1964,Kraichnan1965,Sridhar1994,CriticalBalance,NgBhattacharjee,Revisited1997,Galtier2000,Galtier2006, Mahajan2021}
As an example, turbulence in the solar wind exhibits power law spectra according to the relationship of the magnetic and velocity field for the interacting waves.\cite{MKModeling2005,KMSolar2004,XMHDSolar2016}

The goal of many of the cited turbulence theories is the derivation of energy spectrum (often a power law) for an inertial range in which driving mechanisms balance the effects of dissipation. 
In incompressible magnetohydrodynamics (MHD), for example, Goldreich and Sridhar introduced the critical balance conjecture, which states that the wave spectral content is obtained by balancing the linear oscillation timescale with the nonlinear time scale of wave energy transfer.\cite{CriticalBalance}
This assumption, owing to phenomenological arguments, exploits the eddy-damped quasinormal Markovian (EDQNM) closure to calculate a $k_\perp^{-5/3}$ spectrum.
Another approach to calculation of the spectrum is the Zakharov transformation, which through an assumption of scale invariance in the energy transfer function leads to power law exponents for a stationary spectrum.\cite{Galtier2022}
This has been used by Galtier et al. to calculate a $k_\perp^{-2}$ spectrum for incompressible MHD,\cite{Galtier2000} and a knee in the incompressible Hall MHD spectrum corresponding to where the Hall effect becomes most important.\cite{Galtier2006}

The spatiotemporal chaos which characterizes turbulence also leads to energy content at a range of frequencies. This may be inferred from the spatial distribution of energy from interactions of eddies of similar and disparate scales, straining and sweeping motions in phase space. 
In each case, this leads to a power law frequency spectrum.\cite{Tennekes1975,ChenKraichnan1989,Zhou2004}
Specifically focusing on the linear waves of the system, simulations of compressible Alfvénic turbulence have found a Lorentzian frequency spectrum centered about the Alfvén, slow, and fast modes, with broadening caused by the nonlinear wave interactions.\cite{Yuen2025}
Where low order resonances are difficult to find in a discrete wavenumber system, the phenomenon of nonlinear broadening allows simulations of resonant interactions because resonances exist within the broadening width.\cite{Connaughton2001}
The existence of long time scales of energy transfer is highlighted in the work of Mahajan 2021, which, {\it inter alia}, constructs a reduced model of Hall MHD turbulence by considering a three-wave subset of the total energy spectrum.\cite{Mahajan2021}
This is an exactly integrable system for which a timescale of nonlinear energy transfer may be constructed from the Jacobi elliptic function timescale.

The turbulent correction to linear frequencies has been described in generality by Galtier.\cite{Galtier2022}
This approach uses the method of two-timing, in which the high frequency of a linear system is corrected by longer period behavior.\cite{BenderOrszag}
Such an approach eliminates secular perturbative solutions to the van der Pol and Duffing oscillators,\cite{Strogatz} and is also used by Whitham to analyze nonlinear corrections to a linear dispersion relation for the Klein-Gordon equation.\cite{Whitham1974}
This paper explores a different approach to estimate the nonlinear corrections to linear frequency.
Using the basic framework developed in Mahajan 2021 (to be called Ma21),\cite{Mahajan2021} we will  calculate modified frequencies in a weakly turbulent system by choosing from the slew of nonlinear terms only those that are phase coherent with a given mode. 
These terms nonlinearly modify the linear operator and change the effective frequency via the phase coherent nonlinear terms.

We begin on a similar footing as the multiple scale analysis. We will heavily borrow from Ma21. We begin by reviewing the linear theory and interactions of incompressible Hall MHD in \ref{sec:revhmhd}, concluding with the derivation of the mode coupling coefficients for intra-branch interactions.
At this stage, \ref{sec:coeffs} is devoted to the study of the behavior of these coefficients and in what regions of wavenumber space interactions between waves are strongest.
The break from the multiple-scale method comes in \ref{sec:impnld}, where we invoke  phase-coherent wave interactions to derive a nonlinear dispersion reflecting the effect of perturbations on the linear frequencies.
A calculation for corrections begins in \ref{sec:freq}, where we assume a Maxwellian frequency spectrum for our wave system and use the plasma dispersion function to integrate with a resonant denominator.
The final set of integrations over interacting wave numbers is carried out in \ref{sec:space}, which allows calculation of the nonlinear frequency shifts.
\section{Linear Theory of Incompressible Hall MHD\label{sec:revhmhd}
}

Before calculating of nonlinear frequency shifts, we briefly review the linear waves in an incompressible Hall MHD system immersed in uniform static magnetic field  $B_0 \hat{z}$. We follow here the approach of Ma21. For  an alternative perspective, we refer the interested reader to the work of Galtier in 2006, and Galtier's monograph on MHD in 2016.\cite{Mahajan2021,Galtier2006,Galtier2016}

The perturbed dynamical quantities, the magnetic field $\mathbf{b}$, and the velocity $\mathbf{v}$ are normalized, respectively to  $B_0$ and the the Alfv\'en speed $V_A=B_0/ (4 \pi \rho)^{1/2}$ (with $\rho$ is the constant density). They obey the equations
\begin{align}
    \frac{\partial \mathbf{b}}{\partial t} &= \nabla\times \bigg((\mathbf{v}-(\nabla\times\mathbf{b}))\times (\mathbf{b}+\hat{z})\bigg) \, , \label{eq:beqn} \\
    \frac{\partial \mathbf{v}}{\partial t} &= \mathcal{P}\bigg(\mathbf{v}\times(\nabla\times\mathbf{v})+(\nabla\times\mathbf{b})\times(\mathbf{b}+\hat{z})\bigg)\, , \label{eq:veqn}
\end{align}
where $\mathcal{P}  = \mathcal{I}-\nabla\Delta^{-1}\nabla\cdot$ is a projection operator which reflects the effect of pressure to retain incompressibility. \cite{RoseSulem1978,Chandre2013}
By taking the curl of \eqref{eq:veqn}, we work with the vorticity equation
\begin{align}
    \hspace{-1em} \frac{\partial \nabla\times\mathbf{v}}{\partial t}
    = \nabla\times\bigg(\mathbf{v}\times(\nabla\times\mathbf{v}) + (\nabla\times\mathbf{b})\times(\mathbf{b}+\hat{z})\bigg) \, . \label{eq:vorteqn}
\end{align}
In \eqref{eq:beqn}, \eqref{eq:veqn}, and \eqref{eq:vorteqn}, time scales are set to the ion cyclotron time scale $\omega_{ci}^{-1}$ and length scales to the ion skin depth $d_i = V_A\omega_{ci}^{-1}$.

To derive the linear dispersion, we may readily manipulate the linear system to derive (eliminate $\mathbf{v}$ and using the divergence free conditions),
\begin{align}
    \frac{\partial^2 \nabla\times \mathbf{b}}{\partial t^2} &= \frac{\partial }{\partial z}\bigg(\frac{\partial \nabla\times\mathbf{v}}{\partial t}-\frac{\partial }{\partial t} \nabla\times(\nabla\times\mathbf{b})\bigg) 
    \\ \nonumber
    &= \frac{\partial^2}{\partial z^2 }\nabla\times\mathbf{b} - \frac{\partial}{\partial z}\frac{\partial }{\partial t} \nabla\times(\nabla\times\mathbf{b}) \, 
    \\ \nonumber
    &= \frac{\partial^2}{\partial z^2 }\nabla\times\mathbf{b} + \frac{\partial}{\partial z}\frac{\partial }{\partial t} \nabla^2 \mathbf{b} \, ,
\end{align}
that on Fourier decomposition and rearrangement, yields 
\begin{align}
  i (k_z^2 -\omega^2)\mathbf{k}\times\mathbf{b} = -k_z \omega k^2 \mathbf{b} \, ,
\end{align}
which is the defining equation for a Beltrami field (where the curl of the vector field is aligned with the field). Such a system of circularly polarized waves has been quite extensively studied \cite{Waleffe1992,Mahajan2021} - the salient results are:
\begin{align}
 \mathbf{b} = b \hat{e}_k,  \quad \hat{e}_k = \frac{\mathbf{k}\times\hat{z}}{k_\perp \sqrt{2}} + i \frac{\mathbf{k}\times(\mathbf{k}\times\hat{z})}{kk_\perp \sqrt{2}},
\end{align}
the dispersion relation,
\begin{align}
    \alpha_{k\pm} \equiv \frac{\omega_\pm}{k_z } = \frac{k}{2}\pm \sqrt{1+\frac{k^2}{4}} \label{eq:alphadefn} \, ,
\end{align}
and the proportionality of the perturbations
\begin{align}
\mathbf{v} = -\frac{\mathbf{b}}{\alpha_{k\pm}}\, . \label{eq:bvalpha}
\end{align}
In addition, for the left-handed circularly polarized waves, $i \mathbf{k}\times \hat{e}_k = k \hat{e}_k$.

Notice that the dispersion relation \eqref{eq:alphadefn}  defines two frequency branches for a left-handed circularly polarized wave. We will us the standard nomenclature (Ma21) and refer to $\alpha_{k+}$ as the Alfv\'en whistler-like branch, and $\alpha_{k-}$ as the Alfv\'en cyclotron-like branch.
Repeating this analysis for the right-handed curl eigenstates, for which $\mathbf{b} \propto \hat{e}_k^\dagger$ and $i\mathbf{k}\times \hat{e}_k^\dagger = -k \hat{e}_k^\dagger$, constructs all normal modes of the incompressible Hall MHD system.

\section{Hall MHD Coupling Coefficients \label{sec:coeffs}}

Now that we have applied Fourier analysis to the linearized dynamics, we Fourier transform the nonlinear incompressible Hall MHD equations \eqref{eq:beqn} and \eqref{eq:vorteqn}.
This leads to convolution integrals which determine the evolution of one Fourier component of the magnetic and velocity field perturbations. 
To simplify these expressions, we suppose that the interacting magnetic and velocity field fluctuations are excited along the Alfvén whistler-like branch of the left-handed circularly polarized wave; that is, 
\begin{align}
    \mathbf{b} = b_k \hat{e}_{k} \quad\text{and}\quad \mathbf{v} = -\frac{b_k}{\alpha_{k+}} \hat{e}_k \, .
\end{align}
This restricts our analysis to interactions between whistler-like waves in incompressible Hall MHD.
However, Ma21 suggests that intra-branch interactions are most relevant for the nonlinear dynamics.\cite{Mahajan2021}
Along this branch, we find
\begin{align}
    \frac{db_m}{dt} &- i m_z v_m - i m_z \abs{\mathbf{m}} b_m 
    \\ \nonumber&= \int d^3 n \,
    i \hat{e}_m^\dagger \cdot (\mathbf{m}\times ( \hat{e}_n \times \hat{e}_{m-n}) )
    \\ \nonumber
    \hspace{2cm}&\times  
    (\frac1{\alpha_{n+}}-\abs{\mathbf{m}}) b_n b_{m-n} \, , \\ 
    \frac{dv_m}{dt} &- i  m_z b_m \\ \nonumber
    &= \frac1{\abs{\mathbf{m}}}\int d^3n \, 
    i \hat{e}_m^\dagger \cdot (\mathbf{m}\times (  \hat{e}_n \times \hat{e}_{m-n}))
    \\ \nonumber
    &\quad \quad \times 
    (\frac{\abs{m-n}}{\alpha_{n+} \alpha_{(m-n)+}} + \abs{n})
    b_n b_{m-n}\, .
\end{align}
We rewrite the vector product as 
\begin{align}
    &\hat{e}_m^\dagger \cdot (\mathbf{m}\times(\hat{e}_n \times \hat{e}_{m-n})) \nonumber
    \\ 
    &\quad \quad = (\mathbf{m}\cdot \hat{e}_{m-n})\hat{e}_n\cdot \hat{e}_m^\dagger - (\mathbf{m}\cdot\hat{e}_{n})\hat{e}_{m-n}\cdot \hat{e}_m^\dagger \nonumber \, ,
\end{align}
which prompts the definition of the interaction kernel of the wavenumbers $\mathbf{n}$ and $\mathbf{m}$ to be the product
\begin{align}
    I_{nm} = (\mathbf{m}\cdot \hat{e}_{m-n})(\hat{e}_n \cdot \hat{e}_m^\dagger)
    \label{eq:imn} \, .
\end{align}
By substituting $n\to m-n$ in the integral over the other term in the vector product, all of dependence of the interaction on the orientation of $\mathbf{n}$ and $\mathbf{m}$ is accounted for by $I_{nm}$.
The resulting convolution integrals are 
\begin{align}
    &\int d^3 n \, i g_{nm}I_{nm} b_n b_{m-n}  \, , \\ 
    & \int d^3 n \, i\abs{\mathbf{m}} h_{nm} I_{nm} b_n b_{m-n} \, ,
    \label{eq:finalGH}
\end{align}
where the coefficients $g_{nm}$ and $h_{nm}$ are defined via 
\begin{align}
    g_{nm} &= (\frac1{\alpha_{n+}}-\abs{\mathbf{n}})-(\frac1{\alpha_{(m-n)+}}-\abs{\mathbf{m}-\mathbf{n}}) \, ,\\
    h_{nm} &= \bigg(\frac1{\alpha_{n+}\alpha_{(m-n)+}}-1\bigg)\frac{\abs{\mathbf{m}-\mathbf{n}}-\abs{\mathbf{n}}}{\abs{\mathbf{m}}}\, .
\end{align}
\begin{figure*}[t]
    \centering
    \begin{subfigure}{0.4\linewidth}
        \includegraphics[width=\linewidth]{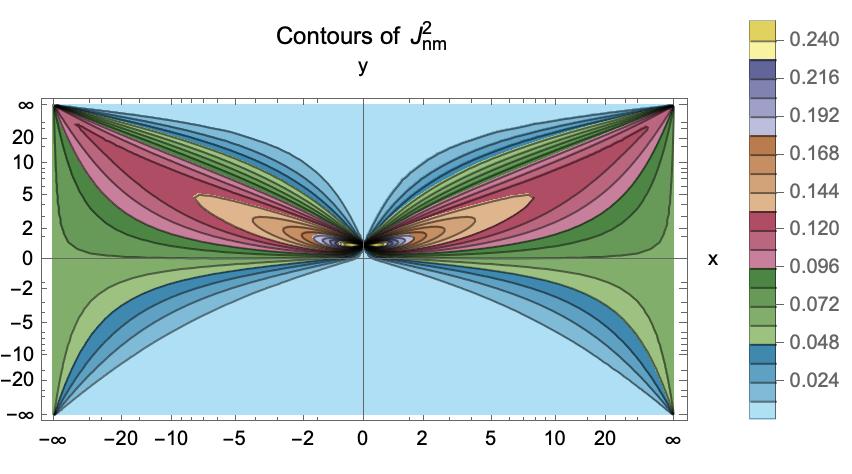}
        \caption{$\abs{\mathbf{m}} = 0.75$}
        \label{fig:jmn2}
    \end{subfigure}
    \hspace{1cm}
    \begin{subfigure}{0.4\linewidth}
        \includegraphics[width=\linewidth]{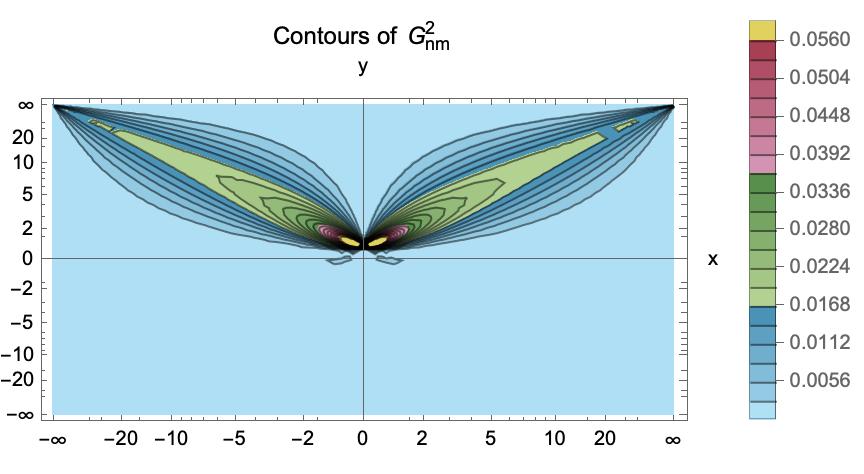}
        \caption{$\abs{\mathbf{m}}=0.65$}
        \label{fig:gmn2}
    \end{subfigure}

    \begin{subfigure}{0.4\linewidth}
        \includegraphics[width=\linewidth]{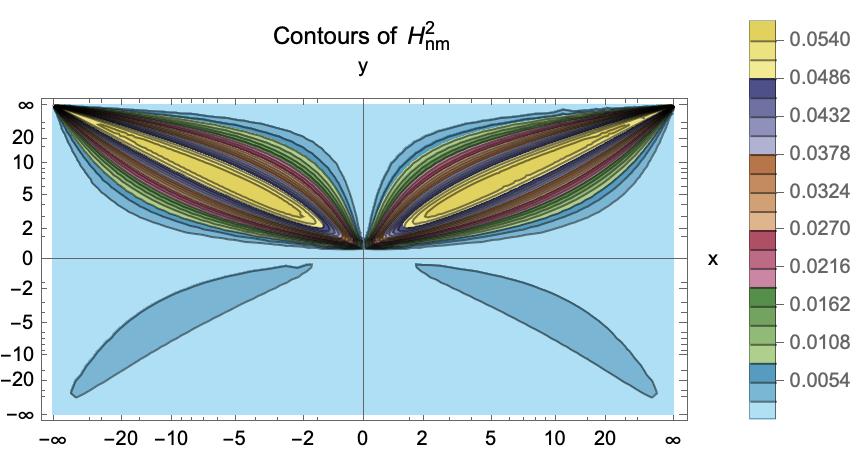}
        \caption{$\abs{\mathbf{m}} = 0.75$}
        \label{fig:hmn2}
    \end{subfigure}
    \hspace{1cm}
    \begin{subfigure}{0.4\linewidth}
        \includegraphics[width=\linewidth]{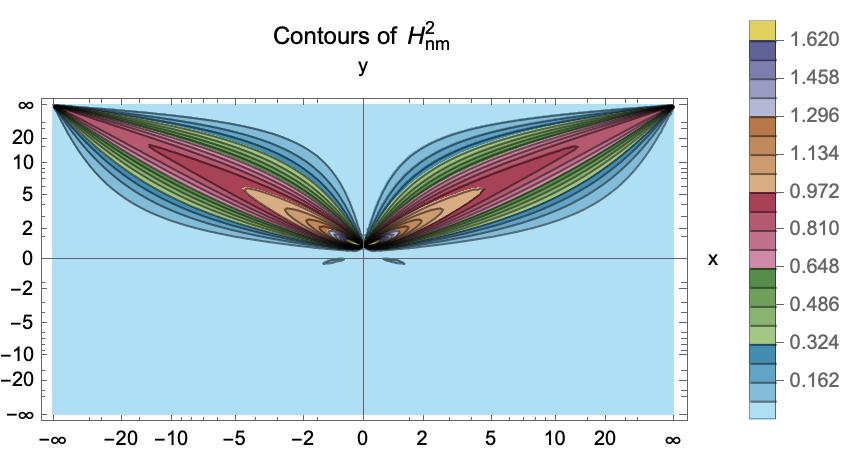}
        \caption{$\abs{\mathbf{m}} = 3$}
        \label{fig:hmn23}
    \end{subfigure}
    \caption{Contours of coupling coefficients of incompressible Hall MHD as a function of $\mathbf{n} = y\, \mathbf{m} + x\, \mathbf{m}\times\hat{z}$.}
    \label{fig:coefcontours}
\end{figure*}

\begin{figure*}[t]
    \centering
    \begin{subfigure}{0.3\linewidth}
        \includegraphics[width=\linewidth]{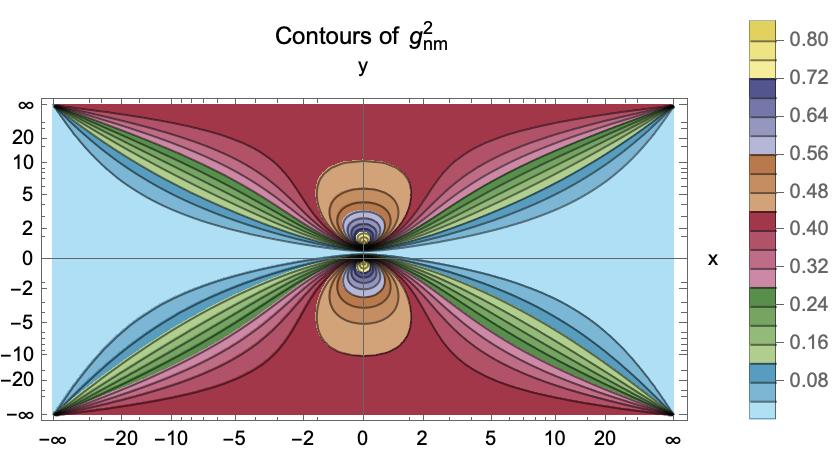}
        \caption{$\abs{\mathbf{m}} = 0.65$}
        \label{fig:lowg}
    \end{subfigure}
    \begin{subfigure}{0.3\linewidth}
        \includegraphics[width=\linewidth]{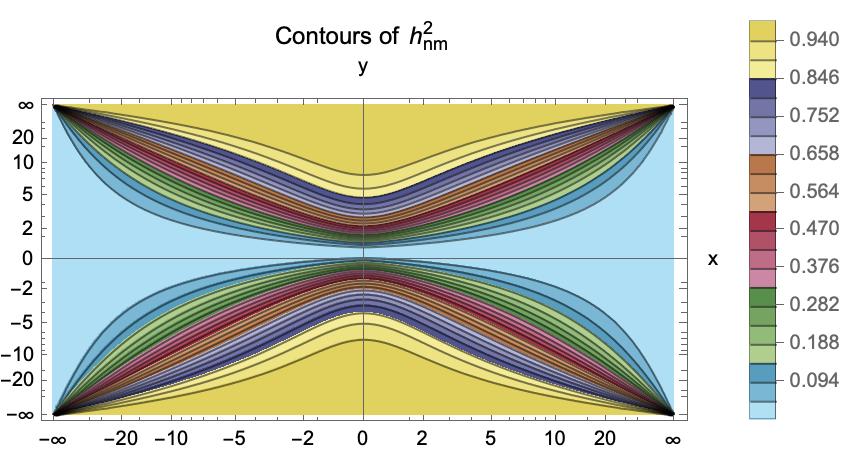}
        \caption{$\abs{\mathbf{m}} = 0.75$}
        \label{fig:lowh1}
    \end{subfigure}
    \begin{subfigure}{0.3\linewidth}
        \includegraphics[width=\linewidth]{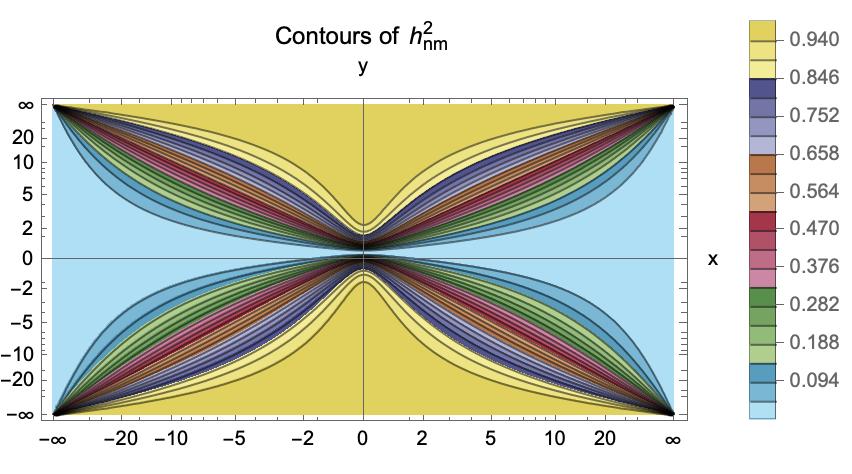}
        \caption{$\abs{\mathbf{m}} = 3$}
        \label{fig:lowh2}
    \end{subfigure}
    \caption{Contours of $g_{nm}$ and $h_{nm}$ corresponding to Figures \ref{fig:gmn2}, \ref{fig:hmn2}, and \ref{fig:hmn23}, respectively.}
    \label{fig:lowcontours}
\end{figure*}

The coefficients $G_{nm}=g_{nm}I_{nm}$ and $H_{nm}=h_{nm}I_{nm}$ calculated above define the energy transfer mechanisms within the Alfv\'en-whistler branch.
To best estimate where interactions are maximized, we analyze these coefficients below.
As a summary, we display contours of $J_{nm}^2 = \abs{I_{nm}}^2$, $G_{nm}^2$, and $H_{nm}^2$ in Figure \ref{fig:coefcontours} and $g_{nm}$ and $h_{nm}$ in Figure \ref{fig:lowcontours}.
These contours are produced by assuming the wavenumber components parallel to $B_0 \hat{z}$ are small and writing the wavenumber $\mathbf{n} = y \, \mathbf{m} + x\, \mathbf{m}\times\hat{z}$.

The interaction kernel appears as a factor determining both $G_{nm}$ and $H_{nm}$, so we begin by analyzing $J_{nm}^2$. This coefficient has a rational function form given by
\begin{align}
    J_{nm}^2 = \frac{\abs{\mathbf{m}}^2}{8} \frac{x^2}{(y-1)^2+x^2}\bigg(1+\frac{y}{\sqrt{y^2+x^2}}\bigg)^2 \, .
\end{align}
The calculation of this form and other useful identities is presented in Appendix \ref{ap:jmn}, and this rational function is displayed in Figure \ref{fig:jmn2}.
This function has no irremovable poles (see next paragraph for some discussion of the removable pole at $\mathbf{n}=\mathbf{m}$), and we observe that $J_{nm}^2$ remains appreciably nonzero for a large range of $\mathbf{n}$.

Despite this, we comment on zeros and maximum values of $J_{nm}^2$.
The fact that $x = 0$ forces $J_{nm}^2 = 0$ suggests that when the wavenumber $\mathbf{n}$ is parallel to $\mathbf{m}$, it does not contribute to its development.
This is a result that has been well established in the literature. \cite{Waleffe1992,Mahajan2021,Galtier2006}
The second factor $1 + y/\sqrt{x^2+y^2}$ approaches zero when $y$ is negative and larger than $x$, which is why the gap in contours is larger around the negative $y$ axis than the positive $y$ axis.
However, there are removable singularities in $J_{nm}^2$ with $\mathbf{n} = \mathbf{m}$.
The root of this lies in the fact that a curl eigenstate $\hat{e}_m$ is undefined when $\mathbf{m}=0$, since at this wavenumber the curl operator is degenerate.

To better understand the functional behavior of $J_{nm}^2$, we provide an alternative interpretation of the removable singularities as a non-existent limit of $J_{nm}^2$.
As described earlier, a limit of $J_{nm}^2$ as $\mathbf{n}\to\mathbf{m}$ with $\mathbf{n}$ parallel to $\mathbf{m}$ must be zero.
On the other hand, the limit with $y = 1$ as $x\to 0$ is $\abs{\mathbf{m}}^2 /{2}$, as the factors of $x^2$ cancel.
This limit is the approach of $\mathbf{n}$ to $\mathbf{m}$ with $\mathbf{n}-\mathbf{m}$ perpendicular to $\mathbf{m}$.
In fact, the bounds on the rational functions comprising $J_{nm}^2$ shows this to be the coefficient's supremum.

Based on the plot of $J_{nm}^2$, we see that if we were to pick a wavenumber about which to expand nonlinear interactions, the perpendicular regions to $\mathbf{n} = \mathbf{m}$ appear most interesting.
However, the slow decline and asymptote along the upper half plane $y = \pm x$ diagonals suggests that there isn't one particular set of interactions that dominates the nonlinear dynamics.

The coupling coefficients themselves $G_{nm}$ and $H_{nm}$ inherit the strength of $J_{nm}^2$ along the upper diagonals. Despite this, the coefficients appear to be maximized by different wavenumbers $\mathbf{n}$.
This is a key feature of the coefficients $\alpha_{n+}$, as the factors of $\abs{\mathbf{m}}$ do not scale out as they do for $J_{nm}^2$. 
We analyze this situation in detail with reference to the prefactor coefficients $g_{nm}$ and $h_{nm}$, displayed in Figure \ref{fig:lowcontours}.
To ease notation, we refer to $\alpha_{k}$ in place of $\alpha_{k+}$.
The equation defining $\alpha_k$ is 
\begin{align}
    \alpha_k^2 - \abs{\mathbf k}\alpha_k - 1 = 0 , 
\end{align}
which implies that $\abs{\mathbf n} = \alpha_n -\alpha_n^{-1}$, so we write $g_{nm}$ and $h_{nm}$ exclusively in terms of $\alpha_n$ and $\alpha_{m-n}$ as
\begin{align}
    g_{nm} &= -(\alpha_n - \alpha_{m-n})(1+\frac{2}{\alpha_n\alpha_{m-n}}) \, ,\\ 
    \abs{\mathbf{m}} h_{nm} &= (\alpha_n - \alpha_{m-n})(1-\frac1{\alpha_n^2\alpha_{m-n}^2}) \, .
\end{align}
The common prefactor of $\alpha_n - \alpha_{m-n}$ reflects the antisymmetry used to isolate the interaction kernel in \eqref{eq:imn}.
It causes $G_{nm}$ and $H_{nm}$ to both vanish when $y = \frac12$.

More interestingly, it offers a unique approach to analyze the behavior of $g_{nm}$ and $h_{nm}$.
Because of the fact that $\mathbf{n}+(\mathbf{m}-\mathbf{n})=\mathbf{m}$, we can construct an upper bound on $\alpha_n - \alpha_{m-n}$ corresponding to the reverse triangle inequality $\abs{\abs{\mathbf n} - \abs{\mathbf m- \mathbf n}} \leq \abs{\mathbf m}$.
In particular, we find
\begin{align}
    &\abs{\alpha_n - \alpha_{m-n} + \alpha_{m-n}^{-1}-\alpha_n^{-1}} 
    \\ \nonumber
    &\hspace{1cm}= (1+\alpha_n^{-1}\alpha_{m-n}^{-1})\abs{\alpha_n - \alpha_{m-n}} \leq \abs{\mathbf{m}} \, .
\end{align}
This gives us bounds on $g_{nm}$ and $h_{nm}$
\begin{align}
    \abs{g_{nm}} \leq &\bigg(1+\frac1{\alpha_n\alpha_{m-n}+1}\bigg)\abs{\mathbf{m}} \, , \\
    \abs{h_{nm}} \leq &\bigg(1 - \frac1{\alpha_n \alpha_{m-n}}\bigg) \, .
\end{align}
We see that the prefactor $g_{nm}$ exceeds its large wavenumber limit $\abs{\mathbf m}$ when one of the two vectors $\mathbf{n}$ or $\mathbf{m-n}$ is small.
The magnitude of this maximum, shown near $(0,0)$ or $(0,1)$ in \ref{fig:lowg}, depends on the size of $\abs{\mathbf m}$, but the maximum value occurs near the removable singularities of $J_{nm}$.
By contrast, $h_{nm}$ is always less than one, and only reaches this value for large, parallel $\mathbf{n}$ and $\mathbf{m-n}$.
It increases towards one more quickly when $\abs{\mathbf m}$ is large so that $1/\alpha_n\alpha_{m-n}$ tends to zero faster, as shown in Figures \ref{fig:lowh1} and \ref{fig:lowh2}.
When this increase is fast enough, $H_{nm}$ remains large at $(0,1)$ and the largest value of $H_{nm}$ also occurs near $(0,1)$.
Otherwise, the product of $h_{nm}$ and $J_{nm}$ is too small close to $\mathbf{n}=\mathbf{m}$, and the value of $\mathbf{n}$ which maximizes $H_{nm}$ occurs deeper into the quadrant as in Figure \ref{fig:hmn2}.
\section{Implicit Nonlinear Dispersion Relation\label{sec:impnld}} 

With the mode coupling coefficients found in the previous section, we are ready to obtain a nonlinear contribution to the dispersion relation.
This process begins by obtaining the linear dispersion relation again, but retaining the nonlinear terms.
To do this, we assume that all interacting waves are circularly polarized so that $\mathbf{b}_m = b_m \hat{e}_m$ and $\mathbf{v}_m = v_m \hat{e}_m$, and we begin from 
\begin{align}
\!\! \frac{db_m}{dt}
    - i m_z v_m  - i m_z \abs{\mathbf{m}} b_m 
   &= \int d^3 n \, i G_{nm} b_n b_{m-n}  \, , \label{eq:bstartdisp} \\
    \frac{dv_m}{dt}
    - i m_z b_m     
    &= \int d^3 n \, i H_{nm} b_n b_{m-n} \, .    
\end{align}
Now we can take the time derivative of \eqref{eq:bstartdisp} to eliminate $\frac{dv_m }{dt}$ as before.
The result is 
\begin{align}
    &\frac{d^2 b_m}{dt^2} - i m_z \abs{\mathbf{m}} \frac{db}{dt} + m_z^2 b_m \\ \nonumber
    &\quad  = \int d^3n \, i \bigg(G_{nm} \frac{d}{dt}(b_n b_{m-n}) + i m_z H_{nm} b_n b_{m-n}\bigg) \, .
\end{align}
This is the linear equation we had solved previously for interacting circularly polarized (whistler-branch) waves.
We may now find nonlinear effects on the frequency by taking a temporal Fourier transform of this equation.
Introducing the symbol $b_{m\omega}$ for $\int b_m(t) e^{-i \omega t }dt$, multiplying by $e^{- i \omega t}$ and integrating yields a nonlinear dispersion relation 
\begin{align}
 & D_{m\omega} b_{m\omega}\equiv (-\omega^2 + m_z \abs{\mathbf{m}} \omega + m_z^2) b_{m\omega} \\ \nonumber
 &\quad = -\int d^3 n  \, (\omega G_{nm} + m_z H_{nm}) \int d\omega' \, b_{n\omega'}b_{(m-n)(\omega-\omega')} \, . \label{eq:nldisp1}
\end{align}
Here $D_{m\omega} = -(\omega-m_z \alpha_{m+})(\omega-m_z \alpha_{m-})$ is an abbreviation for the linear dispersion relation of the system.

We want to manipulate this equation such that it is possible to input a spectral and frequency distribution of $b_{m}$ and return a nonlinear correction to the dispersion relation.
To do this, we can use \eqref{eq:nldisp1} to equally well solve for $b_{n\omega'}$ by writing 
\begin{align}
    b_{n\omega'} = -\frac1{D_{n\omega'}}\int d^3 n' d q \, &(\omega'G_{n'n}+n_z H_{n'n})  \\ \nonumber &\times b_{n'q}b_{(n-n')(\omega'-q)}\, . 
\end{align}
In this convolution integral, we have a term containing $b_{m\omega}$ and $b_{(m-n)(\omega'-\omega)}$, which is phase coherent with our original expansion.
As a result, this term is expected to contribute much more to the development of $b_{m\omega}$ than the random phases present in other modes.
So we approximate
\begin{align}
    b_{n\omega'} = -\frac{\omega'G_{mn}+n_z H_{mn}}{D_{n\omega'}} b_{m\omega}b_{(n-m)(\omega'-\omega)} 
\end{align}
and the evolution of $b_{m\omega}$ is now given by
\begin{align}
    D_{m\omega}b_{m\omega}= -N_{m\omega } b_{m\omega } \, ,
\end{align}
where we have written the nonlinear correction to the linear dispersion relation as $N_{m\omega}$ given by
\begin{align}
    \int d^3 n d\omega' &\frac{(\omega G_{nm}+m_z H_{nm})(\omega' G_{mn}+n_z H_{mn})}{(\omega' -n_z \alpha_{n+})(\omega'-n_z\alpha_{n-})} \\ \nonumber
    &\times \abs{b_{(m-n)(\omega-\omega')}}^2 \, . \label{eq:nldisp}
\end{align}

The following sections will evaluate $N_{m\omega}$ in detail to determine the precise form the frequency shifts.
Before we proceed with this calculation, we show how to find the shift.
We expand the linear frequency as $\omega= \omega_{m+}(1+\zeta)$, where $\omega_{m+}$ satisfies the linear dispersion relation and $\zeta$ is assumed to be $O(E)$, where $E$ is the energy of the magnetic perturbations and should be considered small.
Notice that $N_{m\omega}$ contains the perturbed magnetic energy spectrum, so it is $O(E)$.
To first order in $\zeta$, this equation then reads
\begin{align}
    -\frac{dD_{m\omega}}{d\omega}|_{\omega_{m+}} \omega_{m+} \zeta = N_{m\omega_{m+}} \,. 
\end{align}
The derivative is given by
\begin{align}
    -\frac{dD_{m\omega}}{d\omega}|_{\omega_{m+}} &= 2\omega_{m+}-2 \abs{m} m_z   \\ \nonumber
    &= \frac{2(\omega_{m+}^2- \abs{m}m_z \omega_{m+})}{\omega_{m+}} = \frac{2m_z}{\alpha_{m+}} \, ,
\end{align}
so our generic formula for the nonlinear frequency shift is
\begin{align}
    \omega_{m+}\zeta = \frac{\alpha_{m+}}{2m_z}N_{m\omega_{m+}}\, .
\end{align}
\section{The Frequency Integral\label{sec:freq}}

The nonlinear correction to the dispersion relation calculated in Section \ref{sec:impnld} may now be evaluated for a particular spectral distribution of $b_{(m-n)(\omega-\omega')}$.
This requires an integration over the wavenumbers $\mathbf{n}$ and the frequencies $\omega'$.
The dependence of the nonlinear dispersion relation on $\omega'$ can be reduced to the integrals over rational functions
\begin{align}
    I_{\omega'1} &= \int \frac{\omega' \abs{b_{(m-n)(\omega-\omega')}}^2}{(\omega' -n_z \alpha_{n+})(\omega' - n_z \alpha_{n-})}d\omega'\, , \\ \nonumber
    I_{\omega'2}&= \int \frac{\abs{b_{(m-n)(\omega-\omega')}}^2}{(\omega' -n_z \alpha_{n+})(\omega' - n_z \alpha_{n-})}d\omega' \, ,
\end{align}
which then yield the nonlinear correction
\begin{align}
    N_{m\omega} &=\int d^3n \, (\omega G_{nm}+m_z H_{nm}) \\ \nonumber
     &\quad \quad\times ( G_{mn}I_{\omega'1}+n_z H_{mn} I_{\omega'2}) \,.
\end{align}
We must first impose assumptions how the spectrum $\abs{b_{(m-n)(\omega-\omega')}}^2$ depends on $\omega-\omega'$.
Given that the many frequencies comprising the spectrum have random phases, we assume that the frequency spectrum is Gaussian distributed,\begin{align}
    \abs{b_{(m-n)(\omega-\omega')}}^2 = \frac{\abs{b_{m-n}}^2}{\pi \Delta } e^{-(\frac{\omega-\omega'}{\Delta})^2} \, .
\end{align}
Further applying partial fraction decomposition transforms these integrals to 
\begin{align}
    I_{\omega'1} = &\frac{\abs{b_{m-n}}^2}{\Delta} \int \frac{Ae^{-(\frac{\omega-\omega'}{\Delta})^2}}{\omega'-n_z \alpha_{n+}} - \frac{Be^{-(\frac{\omega-\omega'}{\Delta})^2}}{\omega'-n_z \alpha_{n-}}d\omega' \, , \\ \nonumber    
    I_{\omega'2} = &\frac{\abs{b_{m-n}}^2}{\Delta } \int \frac{Ce^{-(\frac{\omega-\omega'}{\Delta})^2}}{\omega'-n_z \alpha_{n+}} - \frac{Ce^{-(\frac{\omega-\omega'}{\Delta})^2}}{\omega'-n_z \alpha_{n-}}d\omega' \,,
\end{align}
where the coefficients $A$, $B$, and $C$ are given by 
\begin{align}
\hspace{-1em} C  =\frac{1}{n_z (\alpha_{n+}-\alpha_{n-})}\, , A = n_z \alpha_{n+}C\, , B = n_z \alpha_{n-}C \,.
\end{align}

These frequency integrals closely resemble the plasma dispersion function, \cite{FriedConte}
\begin{align}
    Z(\xi) = \frac1{\sqrt{\pi}}\int dx \, \frac{e^{-x^2}}{x-\xi } \, ,
\end{align}
By making the transformation $\omega' = \omega + \Delta x$, the frequency dependence is fully accounted for as
\begin{align}
    \frac{I_{\omega'1}}{\abs{b_{m-n}}^2 \sqrt{\pi }}&= \frac{A Z(\frac{n_z \alpha_{n+}-\omega }{\Delta})-BZ(\frac{n_z \alpha_{n-}-\omega}{\Delta})}{\Delta}\, , \\ \nonumber
    \frac{I_{\omega'2}}{\abs{b_{m-n}}^2 \sqrt{\pi }}&= \frac{C Z(\frac{n_z \alpha_{n+}-\omega}{\Delta})-CZ(\frac{n_z \alpha_{n-}-\omega}{\Delta})}{\Delta}\, .
\end{align}

Before we proceed further, we discuss the limits where the width of the Gaussian distribution $\Delta $ is large and small.
In the white noise limit where $\Delta$ is large, all frequencies are present, and the arguments of the plasma dispersion function are small.
Then the plasma dispersion function behaves as \cite{FriedConte}
\begin{align}
    Z(\xi) = i\sqrt{\pi}e^{-\xi^2} -2\xi + \frac{4}{3}\xi^3 +\ldots \, ,
\end{align}
and the imaginary exponential term dominates at order $O(1)$ (representing the contributions of the poles of the frequency integral at $n_z \alpha_{n\pm}$).
To leading order, $Z(\xi) = i\sqrt{\pi}$ and the frequency integrals become
\begin{align}
    I_{\omega'1} &= i \pi \frac{A-B}{\Delta} \abs{b_{m-n}}^2  = \frac{i \pi\abs{b_{m-n}}^2}{\Delta } \, , \label{eq:freq1low} \\ 
    I_{\omega'2} &= 0 \, ,
    \label{eq:freq2low}
\end{align}
where we used the fact that $A-B = 1$ is an identity from the partial fraction decomposition.

On the other hand, when $\Delta $ is small, in general the arguments of the plasma dispersion function are large and we use the asymptotic form \cite{FriedConte}
\begin{align}
    Z(\xi) = i\sqrt{\pi }e^{-\xi^2} - \frac1\xi -\frac1{2\xi^3} + \ldots \, .
\end{align}
With the coming integration over $\mathbf{n}$, we know that we will find a wavenumber where $\omega' = \omega_{m+}$, picking up the poles from the denominator and requiring the $O(1)$ complex term.
Furthermore, if the frequency is narrow enough, $\omega_{m+}-\omega_{m-} \gg \Delta $, and the second plasma dispersion function will disappear.
Under that consideration, in this limit the integration becomes
\begin{align}
I_{\omega' 1} &= \frac{i \pi A \abs{b_{m-n}}^2}{\Delta} e^{-(n_z \alpha_{n+}-\omega)^2 / \Delta^2 } \, ,\\
I_{\omega' 2} &= \frac{i \pi C \abs{b_{m-n}}^2}{\Delta} e^{-(n_z \alpha_{n+}-\omega)^2 / \Delta^2 } \, .\nonumber  
\end{align}
The factor $1/\Delta$ demonstrates that the narrow frequency spectrum can generate a larger impact on the linear frequencies.
Yet the fast rate of exponential decay suggests that only $\mathbf n$ with $\abs{\mathbf n} = \abs{\mathbf m}$ contribute to the development of $b_m$ in this regime.

The behavior of the plasma dispersion function suggests that the resonant denominators in our nonlinear dispersion relation cause the linear frequencies to become complex.
Consequently, the leading order effect is the growth or dissipation of modes in the Hall MHD system.
Had we expanded to higher order in $E$ in Section \ref{sec:impnld}, oscillation frequency shifts would emerge.

\begin{figure*}[t]
    \centering
    \includegraphics[width=0.8\linewidth]{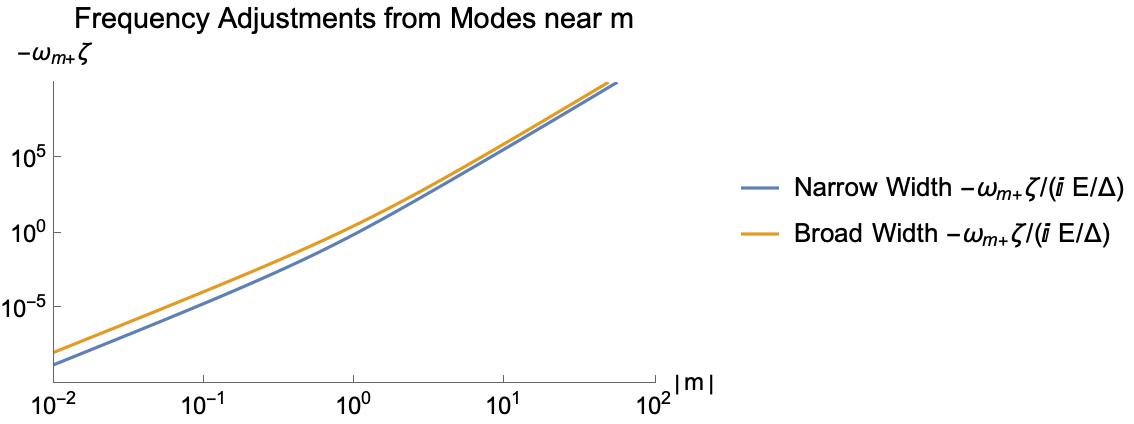}
    \caption{Frequency shift due to wavenumbers near $\mathbf{m}$. We observe what appears to be power law behavior for $\abs{\mathbf{m}}$ large and small, with the frequency shift scaling as $\abs{\mathbf{m}}_\perp^{4}$ for small $\abs{\mathbf{m}}$ and $\abs{\mathbf{m}}_\perp ^6$ for large $\abs{\mathbf{m}}$.}
    \label{fig:narrowfromm}
\end{figure*}

\section{Example Nonlinear Frequency Calculations\label{sec:space}}

In the previous section, we found two limiting forms of nonlinear corrections to the dispersion relation for study of a Gaussian-distributed frequency spectrum.
Both cases give priority to the resonant denominator at $m_z \alpha_{m+}$, which recovers the largest contribution to the frequency shift.
To summarize, we have found for a wide frequency spectrum 
\begin{align}
    N_{m\omega} &=\int d^3n \, (\omega G_{nm}+m_z H_{nm})G_{mn} \frac{i \pi \abs{b_{m-n}}^2}{\Delta } \, , \label{eq:nlcorrhighdelta}
\end{align}
and for a narrow spectrum 
\begin{align}
    N_{m\omega} &=\int d^3n \, (\omega G_{nm}+m_z H_{nm})e^{-(n_z \alpha_{n+}-\omega)^2 / \Delta^2 } \\ \nonumber
     &\quad \quad\times ( n_z \alpha_{n+}G_{mn}+n_z H_{mn} )\frac{i \pi C \abs{b_{m-n}}^2}{\Delta}  \,.\label{eq:nlcorrlowdelta}
\end{align}
Now we must finish the calculation of the frequency shift by integrating over wavenumbers.

We begin the integration over wavenumbers by determining what to do with the integration over the small value $m_z-n_z$.
In the white noise limit \eqref{eq:nlcorrhighdelta},
the only dependence on $n_z$ is in its weak contributions to $G_{mn}$ and $H_{mn}$ through $\abs{n}$ and $\alpha_{n\pm}$.
As a result, if the spectrum is square-integrable and normalized in $n_z$, we may trivially integrate.

But there is explicit dependence on $n_z $ in the delta function frequency limit.
Since the value of $n_z$ must be small to be consistent with our earlier analysis, we assume the spectrum is a sharply peaked Maxwellian around $m_z-n_z = 0$. 
This is well approximated by $\delta(m_z - n_z)$, and the integration trivially replaces $n_z $ by $m_z$.

This enables us to focus the rest of this section on how to evaluate the remaining spatial integrals.
We do this for two examples of frequency spectra.
Because the narrow frequency spectrum rapidly depletes modes away from the circle $\abs{\mathbf n}_\perp =\abs{\mathbf m }_\perp$, we begin by considering a spectrum with $\mathbf{m}\approx \mathbf{n}$.

\subsection{$\mathbf{n}\approx \mathbf{m}$ \label{sec:neqm}}

We assume a spectrum in a cylindrical region centered about the mode $\mathbf{m}$.
For the narrow spectrum, $\Delta$ is small and the spectrum is taken as  
\begin{align}
    b_{m-n}^2 = \begin{cases}
       b & \abs{\mathbf{m}-\mathbf{n}} < \Delta/m_z  \\ \nonumber
       0 & \text{else}
    \end{cases} \,.
\end{align}
For reference, we note in this case the total magnetic energy of the perturbation near $\mathbf m$ is
\begin{align}
E = \pi (\Delta/m_z)^2 b^2 \, .
\end{align}

We carry out the integration for $N_{m\omega_{m+}}$ in Appendix \ref{ap:circle}. The result is 
\begin{align}
    N_{m\omega}=  -\frac{iE}{\Delta}\frac{1.4m_z \abs{\mathbf{m}}^2}{2\alpha_{m+}(1+\alpha_{m+}^{-2})}(\alpha_{m+}g_{mm}+h_{mm})^2 \, ,
\end{align}
which yields a frequency shift
\begin{align}
    -\frac{i E}{\Delta} \frac{1.4}{4}\abs{\mathbf{m}}^2 \frac{(\alpha_{m+}g_{mm}+h_{mm})^2}{(1+1/\alpha_{m+}^2)} \, .
\end{align}

We briefly point out that we can calculate the frequency shift for the large width frequency spectrum for modes close to $\mathbf{m}$.
In this case, $\Delta$ is large and the coefficients will vary considerably on that scale, so we must define a separate scale $\epsilon/m_z$ for the small region of integration.
However, taking similar assumptions leads to the imaginary frequency shift
\begin{align}
    \omega_{m+}\zeta = -\frac{iE}{\Delta} \frac{\pi}{4}\abs{\mathbf{m}}^2 (\alpha_{m+}g_{mm}+h_{mm})\alpha_{m+}g_{mm} \, .
\end{align}

We display the behavior of these frequency shifts in Figure \ref{fig:narrowfromm}.
Despite some differences in their functional form, the corrections appear to have the same power law behavior. 
As we observed in Section \ref{sec:coeffs}, in the large $\abs{\mathbf{m}}$ limit, $h_{mm}\to 1$ where $g_{mm}\to \abs{\mathbf{m}}$, so this owes to the common factors of $\abs{\mathbf{m}}^2\alpha_m^2 g_{mm}^2$.
Moreover, we notice that the numerical factors preceding the coefficients are similar.
As a result, the broad spectrum frequency shift will in practice be much smaller for a comparable energy due to the factor of $\Delta^{-1}$.

This corrective frequency induces an estimate for the nonlinear time scale of energy transfer given by
\begin{align}
    \tau_{nl} = \frac{\Delta }{E }\frac{4/\pi}{\abs{\mathbf{m}}^2 (\alpha_{m+}g_{mm}+h_{mm})\alpha_{m+}g_{mm}} \, .
\end{align}
Notice that this timescale is inversely proportional to the energy of modes concentrated at $\mathbf{m}$, as we could expect that a weak nonlinearity takes longer to modify the dynamics of the system.
As the magnitude of the wavenumber $\abs{\mathbf{m}}$ increases, the timescale for energy transfer gets smaller and can approach the linear time.

\subsection{Leading Order Broad Spectrum \label{sec:genbroad}}

\begin{figure*}

    \centering
    \begin{tikzpicture}[scale=0.7]
        \draw[color=black] (-1,0) -- (6,0);
        \draw[color=black] (0,6) -- (0,-1);
        \filldraw[color=cyan] (0,1) -- (1,0) -- (4.5,3.5) -- (3.5,4.5) -- (0,1);
        \filldraw[color=red] (4.5,3.5) -- (6,5) -- (6,6) -- (5,6) -- (3.5,4.5) -- (4.5,3.5);
        \draw[color=black,very thick] (5,6)-- (0,1) -- (1,0) -- (6,5);
        \draw[color=black,very thick] (4,0) -- (4,6);
        \draw[color=black,very thick] (0,4) -- (6,4);
        \draw[color=black,dashed,very thick] (3.5,4.5) -- (4.5,3.5);
        \path (1,-0.5) node {$\abs{\mathbf{m}}$};
        \path (-0.5,1) node {$\abs{\mathbf{m}}$};
        \path (4,-0.5) node {1};
        \path (-0.5,4) node {1};
        \path (6.5,0) node {$\abs{\mathbf n}$};
        \path (0,6.5) node {$\abs{\mathbf m -\mathbf n}$};
% --- Legend matrix ---
    \node [matrix,draw=black,very thick,row sep=3mm,column sep=3mm] (my matrix) at (9.5,3)
  {
     \node[inner sep=0pt] {\tikz\fill[cyan] (0,0) rectangle (4mm,3mm);}; & \node {$\abs{\mathbf{n}} \ll 1,\ \abs{\mathbf{m}-\mathbf{n}} \ll 1$}; \\ 
     \node[inner sep=0pt] {\tikz\fill[red] (0,0) rectangle (4mm,3mm);}; & \node {$\abs{\mathbf{n}} \gg 1,\ \abs{\mathbf{m}-\mathbf{n}} \gg 1$}; \\
     \node[inner sep=0pt] {\tikz\draw[dashed,very thick] (0,0) -- (3mm,0);}; & \node {$\Sigma =2$}; \\
  };

    \end{tikzpicture}

    \caption{Integration domain in $\abs{\mathbf{n}}$ and $\abs{\mathbf{m}-\mathbf{n}}$ coordinates for small $\abs{\mathbf m}$. We show the regions where $\abs{\mathbf n}$ and $\abs{\mathbf m -\mathbf n}$ are considered large and small, corresponding to where $\Sigma \equiv \abs{\mathbf n} + \abs{\mathbf m -\mathbf n} = 2$.}
    \label{fig:domain}

\end{figure*}
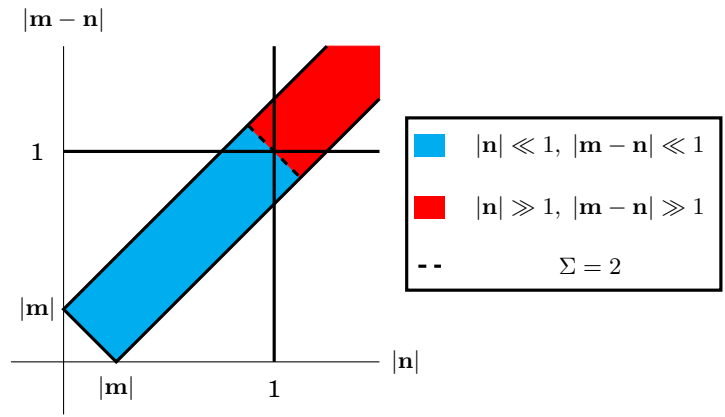

For the narrow frequency spectrum, the exponential damping from the plasma dispersion function concentrates interactions near the circle $\abs{\mathbf m}_\perp = \abs{\mathbf n}_\perp $.
So in this case, we expect interactions with modes near $\mathbf{m}$ to dominate the nonlinear dynamics.
But the broad frequency spectrum places no such constraints on the nonlinear dynamics.
Moreover, the coefficient $g_{mn}$ is qualitatively different from the coefficients $g_{nm}$ and $h_{nm}$ away from the circle $\abs{\mathbf n} = \abs{\mathbf m}$.
It reflects differences between $\abs{\mathbf m -\mathbf n}$ and $\abs{\mathbf m}$, which are of order $\abs{\mathbf n}$.
This complicates efforts to obtain an expression for the frequency shift in terms of the perturbation energy.

Despite this difficulty, we approach the broad frequency spectrum by integrating the leading order behavior of these coefficients directly.
The desired frequency shift is $\frac{\alpha_{m+}}{2m_z}N_{m\omega_{m+}}$, which for small $\abs{\mathbf m}$ is given by 
\begin{align}
    \frac{i\pi}{2\Delta} \int dn_x dn_y \, ( G_{nm}+H_{nm})G_{mn} b_{m-n}^2\, ,
\end{align}
and for large $\abs{\mathbf m}$ becomes 
\begin{align}
    \frac{i\pi \abs{\mathbf{m}}}{2\Delta} \int dn_x dn_y \, (\abs{\mathbf m} G_{nm}+H_{nm})G_{mn} b_{m-n}^2
\, .
\end{align}
This requires asymptotic expressions for $g_{nm}$ and $h_{nm}$, which are summarized in Appendix \ref{ap:ghnm}.
Because of the dependence of $g_{nm}$ and $h_{nm}$ on the norms $\abs{\mathbf{n}}$ and $\abs{\mathbf{m}-\mathbf{n}}$, we find it convenient to use these coordinates for integration.
We display the region of integration, set by the triangle inequalities, in Figure \ref{fig:domain}.

Notice that when $g_{nm}$ and $h_{nm}$ are expanded, we must use as small or large parameters $\abs{\mathbf{n}}$ or $\abs{\mathbf{m}-\mathbf{n}}$.
Figure \ref{fig:domain} assumes that $\mathbf{m}$ is small. In this case, $\abs{\mathbf{n}}$ and $\abs{\mathbf{m}-\mathbf{n}}$ are of similar magnitude (since they differ by at most $\abs{\mathbf m}$), so the boundary $\Sigma \equiv \abs{\mathbf n} + \abs{\mathbf m -\mathbf n}$ is used to distinguish when these variables are large and small.
When $\abs{\mathbf m}$ is large, we only consider the case where all three variables $\abs{\mathbf m}, \abs{\mathbf n}$, and $\abs{\mathbf m-\mathbf n}$ are large, so there are no ``internal boundaries'' used for expansion.

We carry out the integration for the leading order frequency shifts in Appendix \ref{ap:sumdiff}.
This is accomplished by transforming into coordinates 
\begin{align} 
\Sigma = \abs{\mathbf{n}}+\abs{\mathbf{m}-\mathbf{n}} \quad \text{and}\quad 
    r = \frac{\abs{\mathbf{n}}-\abs{\mathbf{m}-\mathbf{n}}}{\abs{\mathbf{m}}} \, ,
\end{align}
and exploiting the fact that $\abs{\mathbf m} < \Sigma$ to estimate the growth rates.
The result for large $\abs{\mathbf m}$ is given by
\begin{align}
    \frac{\pi^2i }{256\Delta}\abs{\mathbf m }^5 \int_{\abs{\mathbf m}}^\infty \Sigma^2 b_{m-n}^2 d\Sigma \, .
\end{align}
In these coordinates, the perturbation energy is
\begin{align}
    \frac{\pi}{2}\int_{\abs{\mathbf m }}^\infty \Sigma b_{m-n}^2 \, d\Sigma \,
\end{align}
so we may interpret the dependence on $g_{mn}$ as relating the frequency shift to the first moment of the energy spectrum.
The dependence as $\abs{\mathbf m}^5$ is consistent with the $\abs{\mathbf m}^6$ scaling we found for modes near $\mathbf{m}$. (In that case, $\Sigma \approx \abs{\mathbf m}$).

The frequency shift for small $\abs{\mathbf m}$ is given by 
\begin{align}
    &\nonumber -\frac{\pi^2i}{256\Delta} \abs{\mathbf m}^2 \\ \nonumber &\times 
    \bigg(\int_{2}^\infty \Sigma^2 b_{m-n}^2 d\Sigma  
    + \frac34 \int_{\abs{\mathbf m}}^2 \Sigma^3 b_{m-n}^2 \, d\Sigma \,\bigg) \, ,
\end{align}
where the contribution from large scales $\Sigma > 2$ is given at left and from large scales $\Sigma < 1$ is given at right.
Now we find $\abs{\mathbf m}^2$ dependence. 
This is fundamentally lower order than \ref{sec:neqm} and owes to the fact that in general $h_{nm}$ is $O(1)$ where $h_{mm}$ is $O(\abs{\mathbf m})$. 
The additional factors of $\Sigma$ comprising $h_{nm}$ at large scales introduces the second moment of the energy spectrum in the calculation of this frequency shift.
\section{Summary}

Most turbulence problems are essentially unsolvable in entirety. 
But to gain theoretical understand one always deals with a subset of the total problem. 
Such theoretical subproblems are rather important in not only advancing our understanding, but also in interpreting simulations that, at least, attempt to do the whole problem.
The word nonlinear, in this paper, connotes the subset of nonlinearities that tend to ``modify'' the linear dispersion relation through nonlinear terms that are phase coherent with the mode under consideration.

The analysis set forth in his paper, therefore, quantifies the modification of frequencies due to this subset of nonlinear interactions in the Hall MHD system. 
Of particular interest is the investigation of the behavior of the projection of the interaction kernel that reflects the geometric impact of a triadic interaction.
Even for this highly simplified nonlinear interaction, the algebra is rather complicated. 
As no region in wavenumber space strongly sets the interactions, the rational structure of the interaction kernel function complicates the turbulent dynamics.

That a perpendicular approach to the $\mathbf{n}=\mathbf{m}$ point is the supremum of the interaction kernel, yet a parallel approach is the well-known null interaction found by Waleffe, accentuates the detailed dependence of the nonlinear coupling on wavenumber.\cite{Waleffe1992}
This removable singularity is the only place the coupling coefficients are not analytic.
The procedure of this paper can be used in a straightforward manner to find frequency corrections due to interactions at any point in wavenumber space.

The frequency shift we have calculated has direct consequences for the calculation of the energy spectrum. 
The frequency shifts calculated here come out to be, primarily, damping or growth depending on the wave number - measuring the times for energy exchange and rearrangement in modes. 
Using the familiar conjecture of critical balance,\cite{CriticalBalance} i.e. by equating the nonlinear time for energy transfer to the linear timescale of the Alfvén wave, we can find the energy spectrum. Such a prescription for the current calculation of Hall MHD will lead to 
\begin{align}
    E_m \propto m_z \alpha_m \abs{\mathbf{m}}^{-k}
    \, .
\end{align}
a typical power law shape. Given that $k = 4$ for small $\abs{\mathbf{m}}$ and $k=6$ for large $\abs{\mathbf{m}}$, this recovers the known shift in the energy spectrum scaling when length scales pass the ion skin depth.

The analysis of this paper focuses on interactions of the whistler-like branch in Hall MHD.
Cyclotron-like branch wave interactions can be recovered by a similar argument, with the transformation $\alpha_{n-} = -\alpha_{n+}^{-1}$ to find maxima of $G_{nm}$ and $H_{nm}$ on this branch.
Further exploration of the coupling coefficients via the multiple scale method as in the general work of Galtier would be useful to verify the dependence of frequency corrections on the wavenumber in a general case.\cite{Galtier2022}

\appendix
\section{Reduced Form for Interaction Kernel} \label{ap:jmn}

In this section, we derive the reduced form of the magnitude of the interaction kernel used in the main text,
\begin{align}
    J_{nm}^2 = \frac{\abs{\mathbf{m}}^2}{8}\frac{x^2}{(y-1)^2+x^2}\bigg(1+\frac{y}{\sqrt{y^2+x^2}}\bigg)^2 \,,
\end{align}
where the wavenumber $\mathbf{n}$ is taken to be 
\begin{align}
    \mathbf{n} =  y \, \mathbf{m} + x \, \mathbf{m} \times \hat{z} \,.
\end{align}
This incorporates the assumption that the field-aligned components of $\mathbf{m}$ and $\mathbf{n}$ are small, which further implies $\abs{\mathbf{m}}\sim \abs{\mathbf{m}}_\perp$.

The two factors comprising the interaction kernel each require a dot product in terms of the circularly polarized vectors $\hat{e}_m$, $\hat{e}_n$, and $\hat{e}_{m-n}$, which have the representation
\begin{align}
    \hat{e}_{m} = \frac{\mathbf{m}\times\hat{z}}{\abs{\mathbf{m}}_\perp\sqrt{2}} + i \frac{\mathbf{m}\times(\mathbf{m}\times\hat{z})}{\abs{\mathbf{m}}\abs{\mathbf{m}}_\perp\sqrt{2}} \,.
\end{align}
This form can be exploited to write the circularly polarized vectors in the eigenbasis $\mathbf{m},\mathbf{m}\times\hat{z},\mathbf{m}\times (\mathbf{m}\times\hat{z})$.
Note that
\begin{align}
    (\mathbf{m}\times\hat{z})\times\hat{z} = m_z \hat{z} - \mathbf{m} \sim - \mathbf{m} \,
\end{align}
so that $\mathbf{n}\times \hat{z} = y \, \mathbf{m}\times\hat{z} - x \,\mathbf{m}$ and the left handed vector corresponding to $\mathbf{n}$ is
\begin{align}
    \hat{e}_n &= \frac{\mathbf{n}\times\hat{z}}{\abs{\mathbf{n}}_\perp \sqrt{2}} + i \frac{\mathbf{n}\times(\mathbf{n}\times\hat{z})}{\abs{\mathbf{n}}\abs{\mathbf{n}}_\perp \sqrt{2}}
    \\ \nonumber
    &= \frac{y \, \mathbf{m}\times\hat{z} - x\, \mathbf{m}}{\abs{\mathbf{n}}_\perp \sqrt{2} } \\ \nonumber
    &+ i \frac{(y \, \mathbf{m} + x \, \mathbf{m}\times\hat{z})\times (y \, \mathbf{m}\times\hat{z} - x\, \mathbf{m})}{\abs{\mathbf{n}}\abs{\mathbf{n}}_\perp \sqrt{2} } \\ \nonumber
    &= \frac{y \, \mathbf{m}\times\hat{z} - x\, \mathbf{m}}{\abs{\mathbf{n}}_\perp \sqrt{2} }
    + i \frac{(x^2+y^2)\,\mathbf{m}\times(\mathbf{m}\times\hat{z})} {\abs{\mathbf{n}}\abs{\mathbf{n}}_\perp \sqrt{2} }\,.
\end{align}
Given $\mathbf{m}-\mathbf{n} = (1-y)\, \mathbf{m} - x \, \mathbf{m}\times\hat{z}$, the substitution $x\to -x$ and $y\to (1-y)$ shows that
\begin{align}
    \hat{e}_{m-n}&= \frac{(1-y) \, \mathbf{m}\times\hat{z} + x\, \mathbf{m}}{\abs{\mathbf{m-n}}_\perp \sqrt{2} } \\ \nonumber
    &+ i \frac{(x^2+(1-y)^2)\,\mathbf{m}\times(\mathbf{m}\times\hat{z})} {\abs{\mathbf{m-n}}\abs{\mathbf{m-n}}_\perp \sqrt{2} }\,.
\end{align}
This form allows us to calculate the first factor of the interaction kernel as 
\begin{align}
    \mathbf{m}\cdot \hat{e}_{m-n}
    = \frac{x\abs{\mathbf{m}}^2}{\abs{\mathbf{m}-\mathbf{n}}_\perp \sqrt{2}} 
\end{align}
and the second factor as 
\begin{align}
    \hat{e}_m^\dagger \cdot \hat{e}_n &= \frac{\mathbf{m}\times\hat{z}}{\abs{\mathbf{m}}_\perp\sqrt{2}} \cdot \hat{e}_n - i \frac{\mathbf{m}\times(\mathbf{m}\times\hat{z})}{\abs{\mathbf{m}}\abs{\mathbf{m}}_\perp\sqrt{2}} \cdot \hat{e}_n \\ \nonumber
    &= \frac{y \, \abs{\mathbf{m}}_\perp ^2 }{2\abs{\mathbf{m}}_\perp \abs{\mathbf{n}}_\perp } + \frac{(x^2 +y^2)\, \abs{\mathbf{m}\times(\mathbf{m}\times \hat{z})}^2}{2\abs{\mathbf{m}}\abs{\mathbf{m}}_\perp \abs{\mathbf{n}}\abs{\mathbf{n}}_\perp } \, .
\end{align}
We notice that both factors are purely real, which agrees with previous calculations of the interaction kernel for small $m_z$ and $n_z$. \cite{Mahajan2021}
Combining these factors with the result that $\abs{\mathbf{n}}^2 = (y^2 + x^2)\, \abs{m}^2 $, $\abs{\mathbf{m}-\mathbf{n}}^2 = ((1-y)^2 + x^2)\, \abs{m}^2$, and $\abs{\mathbf{m}\times(\mathbf{m}\times\hat{z})} = \abs{\mathbf{m}}\abs{\mathbf{m}}_\perp$, the modulus squared of the interaction kernel is 
\begin{align}
    J_{nm}^2 = \frac{\abs{\mathbf{m}}^2}{8} \frac{x^2 }{(1-y)^2+x^2}\bigg(1+\frac{y}{\sqrt{x^2+y^2}}\bigg)^2 \, . \label{eq:xyjmn2}
\end{align}
We also write the interaction kernel in different coordinate systems prepare for the integration of Section \eqref{sec:space}.
If we take $x=r\sin \theta $ and $y=1-r\cos \theta $ to be the scaled coordinates of the vector $\mathbf{m}-\mathbf{n}$, the interaction kernel becomes 
\begin{align}
\label{polarjmn2}
    J_{nm}^2 = \frac{\abs{\mathbf{m}}^2}{8}
    \sin^2\theta 
    \bigg(1+\frac{1-r\cos\theta}{\sqrt{1+r^2-2r\cos\theta}}\bigg)^2 
    \, . 
\end{align}
This form is of interest because the $\sin^2\theta$ factor reflects the interaction kernel's dependence on the angle of approach of $\mathbf{n}$ to $\mathbf{m}$.
If $\theta = 0$, $\mathbf{n}$ and $\mathbf{m}$ are parallel and there is no interaction.
If $\theta = \pi/2$, $\mathbf{m}-\mathbf{n}$ is perpendicular to $\mathbf{m}$ and the interaction is maximized.

It is also possible to use as coordinates the moduli $\abs{\mathbf{n}}$ and $\abs{\mathbf{m-n}}$.
In this system, we have 
\begin{align}
\hspace{-1em}
    \frac{\abs{\mathbf{n}}^2}{\abs{\mathbf{m}}^2} = x^2 + y^2 \quad\text{and}\quad \frac{\abs{\mathbf{m}-\mathbf{n}}^2}{\abs{\mathbf{m}}^2} = x^2+(y-1)^2\, ,
\end{align}
so that
\begin{align}
    y = \frac12 +\frac{\abs{\mathbf{n}}^2-\abs{\mathbf{m}-\mathbf{n}}^2}{2\abs{\mathbf{m}}^2} \, 
\end{align}
and $x = \sqrt{(\abs{\mathbf{n}}/\abs{\mathbf{m}})^2 - y^2}$. 
Abbreviating $n_y = \abs{\mathbf{m}} y$ leads to the following form of the interaction kernel,
\begin{align}
    J_{nm}^2 = \frac{\abs{\mathbf{m}}^2}{8} \frac{(\abs{\mathbf{n}
    }-n_y)(\abs{\mathbf{n}}+n_y)^3}{\abs{\mathbf{m}-\mathbf{n}}^2 \abs{\mathbf{n}}^2} 
    \label{eq:jnorms} \, . 
\end{align}

We conclude this section by writing the reduced form of the interaction kernel in a coordinate independent way.
Using the fact that  \begin{align}
    x = \frac{(\mathbf m \times \hat{z})\cdot \mathbf n }{\abs{\mathbf m}^2} \quad \text{and} \quad 
    y = \frac{\mathbf m \cdot \mathbf n}{\abs{\mathbf m}^2} \, ,
\end{align}
the interaction kernel becomes
\begin{align}
    I_{nm} = \frac{\hat{z}\cdot(\mathbf n \times \mathbf m )}{2\sqrt{2}\abs{\mathbf m - \mathbf n}}\bigg(1+\frac{\mathbf n \cdot \mathbf m }{\abs{\mathbf n}\abs{\mathbf m}}\bigg)\,. \label{eq:Icoordless}
\end{align}

\section{Integration of a Circular Spectrum \label{ap:circle}}

Our desired nonlinear component of the dispersion relation $N_{m\omega_{m+}}$ is
\begin{align}
    \frac{i \pi  m_z^2C b^2}{\Delta}\int&dn_x dn_y \,  (\alpha_{m+}g_{nm}+h_{nm})\\ \nonumber
    &\times I_{nm}I_{mn} (\alpha_{n+}g_{mn}+h_{mn}) e^{-\frac{(\alpha_{n+}-\alpha_{m+})^2}{\Delta^2 /m_z^2}} \, .
\end{align}
We are going to integrate using a polar coordinate system with $n_x = \abs{\mathbf{m}} r\sin\theta$ and $n_y = \abs{\mathbf{m}} - \abs{\mathbf{m}}r\cos\theta$.
However, we suppose that the region of integration of size $\Delta$ is very small, so that the exponential factor has the fastest variation in $r$ and the coefficients $\alpha_{n+}$, $g_{nm}$, and $h_{nm}$ may be taken as the constants $\alpha_{m+}$, $g_{mm}$, and $h_{mm}$.

Using the coordinate independent form of the interaction kernel \eqref{eq:Icoordless} shows that with small parallel wavenumbers, $I_{mn} = -I_{nm} = -J_{nm}$.
Furthermore, the impact of the removable pole of the interaction kernel is fully reflected by the sine function in \eqref{polarjmn2}, 
\begin{align}
    J_{nm}^2 &= \frac{\abs{\mathbf{m}}^2}{8}\sin^2\theta \bigg(1+\frac{1-r\cos\theta}{\sqrt{1+r^2-2r\cos\theta}}\bigg)^2 \nonumber \\ &\approx \frac{\abs{\mathbf{m}}^2}{2}\sin^2\theta\, .
\end{align}

To estimate the exponential term, we approximate 
\begin{align}
    \nonumber 
    \alpha_{n+}-\alpha_{m+} &\approx \frac{d\alpha_{m+}}{d\abs{\mathbf{m}}} (\abs{\mathbf{n}}-\abs{\mathbf{m}}) \\ \nonumber
    &= \abs{\mathbf{m}}\frac{d\alpha_{m+}}{d\abs{\mathbf{m}}}(\sqrt{1-2r\cos\theta + r^2}-1) \\ \nonumber
    &\approx -\abs{\mathbf{m}}\frac{d\alpha_{m+}}{d\abs{\mathbf{m}}} r\cos\theta \, ,
\end{align}
which brings the exponent to the variables of integration.
Through implicit differentiation of 
\begin{align}
    \nonumber \alpha_{m+}^2 - \abs{\mathbf{m}} \alpha_{m+} - 1=0
\end{align}
we find
\begin{align}
    \frac{d\alpha_{m}}{d\abs{\mathbf{m}}} = \frac{\alpha_{m+}}{2\alpha_{m+}-\abs{\mathbf{m}}} = \frac{1}{1+\alpha_{m+}^{-2}} \,.
\end{align}
This is a slowly varying function that ranges between $1$ for large $\abs{\mathbf{m}}$ and $1/2$ for small $\abs{\mathbf{m}}$.
To make this integration tractable, we assume it is constant and equal to $1/d$.

Rescaling $r$ by by $\sqrt{d} \frac{\Delta}{m_z \abs{\mathbf{m}}}$ culminates in the expression for $N_{m\omega_{m+}}$
\begin{align}
   -&\frac{i \pi m_z^2 C b^2 \abs{\mathbf{m}}^4 }{2\Delta}(\alpha_{m+}g_{mm}+h_{mm})^2 (\frac{\Delta}{m_z \abs{\mathbf{m}}})^2 I \, ,
\end{align}
with an integral $I$ which we evaluate numerically
\begin{align}
    I &=d \int_0^{1/\sqrt{d}} \int_0^{2\pi} r e^{-r^2 \cos^2 \theta}\sin^2\theta \, d\theta dr \\ \nonumber
    &=\begin{cases}
        1.403 & d =1 \\ 
        1.480 & d=2
    \end{cases} \, . 
\end{align}
Given the similarity in these values, we retain the $d = 1$ value and present the result as
\begin{align}
    \hspace{-0.8em} N_{m\omega}=  -\frac{iE}{\Delta}\frac{1.4m_z \abs{\mathbf{m}}^2}{2\alpha_{m+}(1+\alpha_{m+}^{-2})}(\alpha_{m+}g_{mm}+h_{mm})^2 \, . \label{eq:circle}
\end{align}

\section{Leading Order Expansion of $g_{nm}$ and $h_{nm}$ \label{ap:ghnm}}

By using the fact that $\alpha_{m+}^{-1} = - \alpha_{m-}$, we write $g_{nm}$ and $h_{nm}$ in the forms
\begin{align}
    &g_{nm} = \sqrt{1+\frac{\abs{\mathbf{n}}^2}{4}} - \frac32 \abs{\mathbf{n}} - \sqrt{1+\frac{\abs{\mathbf{m}-\mathbf{n}}
    ^2}{4}}+\frac32 \abs{\mathbf{m}-\mathbf{n}} \, , \nonumber \\ \nonumber 
    &h_{nm} =\frac{\abs{\mathbf{m}-\mathbf n}-\abs{\mathbf n}}{\abs{\mathbf m}} \\ \nonumber
    &\times \bigg( (\sqrt{1+\frac{\abs{\mathbf n}^2}{4}} - \frac{\abs{\mathbf n}}{2})(\sqrt{1+\frac{\abs{\mathbf m- \mathbf n}^2}{4}}-\frac{\abs{\mathbf m-\mathbf n}}{2})-1\bigg) \, .
\end{align}
When both $\abs{\mathbf{n}}$ and $\abs{\mathbf{m}-\mathbf{n}}$ are small, the square root terms simplify to one, and at leading order 
\begin{align}
    g_{nm} &\approx -\frac32 (\abs{\mathbf{n}}-\abs{\mathbf{m}-\mathbf{n}}) \, , \\ \nonumber
    h_{nm} &\approx \frac{\abs{\mathbf{m}-\mathbf n}-\abs{\mathbf n}}{\abs{\mathbf m}}\times- \bigg(\frac{\abs{\mathbf n}+\abs{\mathbf m - \mathbf n}}{2}\bigg) \\ \nonumber
    &\approx \frac{\abs{\mathbf n}^2 - \abs{\mathbf m - \mathbf n}^2}{2 \abs{\mathbf m}}\, .
\end{align}
In the opposite limit, $\sqrt{1 + (x/2)^2} \approx x/2$, and 
\begin{align}
    g_{nm} &\approx - (\abs{\mathbf{n}}-\abs{\mathbf{m}-\mathbf{n}}) \, , \\ \nonumber
    h_{nm} &\approx \frac{\abs{\mathbf{n}}-\abs{\mathbf{m}-\mathbf{n}}}{\abs{\mathbf m}} \, .
\end{align}

\section{General Integration of the Broad Frequency Spectrum \label{ap:sumdiff}}

To integrate the broad frequency spectrum, we use the sum and difference coordinates
\begin{align}
    \Sigma = \abs{\mathbf{n}}+\abs{\mathbf{m}-\mathbf{n}} \quad \text{and}\quad 
    r = \frac{\abs{\mathbf{n}}-\abs{\mathbf{m}-\mathbf{n}}}{\abs{\mathbf{m}}} \, ,
\end{align}
which models the domain shown in Figure \ref{fig:domain} to the integration bounds $-1\leq r \leq 1$ and $\abs{\mathbf m } \leq \Sigma \leq \infty$.
We begin the integration process by rewriting $g_{nm},h_{nm},$ and $J_{nm}^2$ in terms of the sum and difference coordinates.

With the leading order expansions of $g_{nm}$ and $h_{nm}$ carried out in Appendix \ref{ap:ghnm}, we easily write down
\begin{align}
    g_{nm} &= \begin{cases}
        -\frac32 r \abs{\mathbf m} & \abs{\mathbf n },\abs{\mathbf m - \mathbf n } \ll 1\\
        - r \abs{\mathbf m} & \abs{\mathbf n },\abs{\mathbf m - \mathbf n } \gg 1
    \end{cases} \, , \\ 
    h_{nm} &= \begin{cases}
        \frac12 r \Sigma  & \abs{\mathbf n },\abs{\mathbf m - \mathbf n } \ll 1\\
        r & \abs{\mathbf n },\abs{\mathbf m - \mathbf n } \gg 1 
    \end{cases}\, .
\end{align}
Notice that $-1\leq r \leq 1$ by the triangle inequality, so $r$ is independent of the expansion parameter $\abs{\mathbf m}$.
As a result, we notice from our leading order forms that $h_{nm}$ is $O(1)$ while $g_{nm}$ is $O(\abs{\mathbf{m}})$.
Since $J_{nm}^2$ is $O(\abs{\mathbf m}^2)$, our general broad spectrum frequency shift returns a small $\abs{\mathbf{m}}$ frequency shift at order $O(\abs{\mathbf m}^3)$.
This is at lower order than the frequency shift previously calculated in Section \ref{sec:neqm}.
The reason for this is that in that case, $\Sigma = \abs{\mathbf{m}}$ so $h_{nm}$ was also $O(\abs{\mathbf m})$.

In general, we expect that $g_{mn}$ will be related to the difference between $\abs{\mathbf m - \mathbf n} $ and $\abs{\mathbf m}$, which is of order $O(\abs{\mathbf n})$.
A similar expansion for $g_{mn}$ reveals that
\begin{align}
    g_{mn} = \begin{cases}
        \frac34 \Sigma (1-(r+2)\frac{\abs{\mathbf m }}{\Sigma}) & \abs{\mathbf m },\abs{\mathbf m - \mathbf n}, \abs{\mathbf n} \ll 1 \, ,\\
        \frac12 \Sigma (1-(r+3)\frac{\abs{\mathbf m }}{\Sigma}) & \abs{\mathbf m }\ll 1,\abs{\mathbf m - \mathbf n}, \abs{\mathbf n} \gg 1\\
        \frac12 \Sigma (1-(r+2)\frac{\abs{\mathbf m }}{\Sigma}) & \abs{\mathbf m },\abs{\mathbf m - \mathbf n}, \abs{\mathbf n} \gg 1 \, ,
    \end{cases} \,, \nonumber
\end{align}
where we include the separate case where $\abs{\mathbf m} \ll 1$ and $\abs{\mathbf n},\abs{\mathbf m - \mathbf n} \gg 1$.
(We also neglect the small region where $\abs{\mathbf m - \mathbf n} \ll 1$ but $\abs{\mathbf m}, \abs{\mathbf n} \gg 1$).

Progressing with the calculation, we note that 
\begin{align}
    2\abs{\mathbf{n}} &= \Sigma + \abs{\mathbf{m}} r \, ,\\ \nonumber
    2\abs{\mathbf{m}-\mathbf{n}} &=\Sigma - \abs{\mathbf{m}} r \, , \\ \nonumber
    2n_y &= \Sigma r+\abs{\mathbf{m}} \, , \\ \nonumber
    2(\abs{\mathbf{n}}\pm n_y) &= (1\pm r) (\Sigma\pm \abs{\mathbf{m}}) \, , 
\end{align}
and substituting this transformation into the interaction kernel \eqref{eq:jnorms} yields 
\begin{align}
    J_{nm}^2 &= \frac{\abs{\mathbf{m}}^2}{8} (1-r)(1+r)^3  
    \\ \nonumber
    &\times \frac{(1-\abs{\mathbf{m}}/\Sigma)(1+\abs{\mathbf{m}}/\Sigma)^3}{(1-\abs{\mathbf{m}}^2r^2/\Sigma^2)^2} \, .
    \label{eq:jsumdiff}
\end{align}

Our last step before integrating is the computation of the Jacobian for the transformation to sum and difference coordinates.
The matrix of partial derivatives with respect to $n_x$ and $n_y$ is 
\begin{align}
    \begin{bmatrix}
    \frac{n_x}{\abs{\mathbf n }}+\frac{n_x }{\abs{\mathbf m - \mathbf n }}
    & 
    \frac{n_y}{\abs{\mathbf n }} + \frac{n_y - \abs{\mathbf m }}{\abs{\mathbf m - \mathbf n }} 
    \\ 
    \frac1{\abs{\mathbf{m}}} (\frac{n_x}{\abs{\mathbf n }}-\frac{n_x }{\abs{\mathbf m - \mathbf n }}) 
    &
    \frac1{\abs{\mathbf{m}}} ( \frac{n_y}{\abs{\mathbf n }} - \frac{n_y - \abs{\mathbf m }}{\abs{\mathbf m - \mathbf n }}) 
    \end{bmatrix} \, .
\end{align}
In calculation of the determinant, only the terms multiplying $\frac{\abs{\mathbf m}}{\abs{\mathbf m - \mathbf n}}$ do not cancel, and the result is
\begin{align}
    \mathcal{J}^{-1} = \frac{2 n_x }{\abs{\mathbf n } \abs{\mathbf m - \mathbf n}} \, .
\end{align}
Using the identities above simplifies the result to
\begin{align}
    \mathcal{J}^{-1} = \frac{2 \sqrt{(1+r)(1-r)(1-\abs{\mathbf{m}}^2 / \Sigma^2)}}{\Sigma (1-\abs{\mathbf m }^2r^2/\Sigma^2)} \, .
\end{align}

To carry out the integration and achieve our main result, we simplify our expressions by assuming that $\abs{\mathbf m}/\Sigma \ll 1$.
This is justified because $\abs{\mathbf m} < \Sigma$ in any case, and inaccuracies are only appreciable within a few multiples of $\abs{\mathbf m}$.
Such an assumption also reduces the wavenumber dependence of the energy spectrum to dependence on $\Sigma$.

Now we are prepared to integrate leading order contributions.
For large $\abs{\mathbf m}$, the frequency shift is given by
\begin{align}
    -\frac{i\pi \abs{\mathbf{m}}}{2\Delta} \int dn_x dn_y \, \abs{\mathbf m} g_{nm}g_{mn}J_{nm}^2 b_{m-n}^2
\end{align}
where we have discarded the $h_{nm}$ contribution because it is $O(1)$ while the $g_{nm}$ contribution is $O(\abs{\mathbf m})$.
The simplification above leads to the frequency shift
\begin{align}
\hspace{-1em}    \frac{i\pi \abs{\mathbf{m}}^5}{64\Delta} \int_{\abs{\mathbf m}}^\infty  
    d\Sigma \, \Sigma^2 b_{m-n}^2 \int_{-1}^1  dr \, \frac{r(1-r)(r+1)^3}{\sqrt{1-r^2}} \, .
\end{align}
We evaluate the integration over $r$ as 
\begin{align}
    \int_{-1}^1 \frac{r(1-r)(1+r)^3}{(1+r)^{0.5}(1-r)^{0.5}} dr = \frac{\pi}{4} \, , 
\end{align}
simplifying the large $\abs{\mathbf m}$ frequency shift to 
\begin{align}
    \frac{\pi^2i }{256\Delta}\abs{\mathbf m }^5 \int_{\abs{\mathbf m}}^\infty \Sigma^2 b_{m-n}^2 d\Sigma \, .
\end{align}

For the small $\abs{\mathbf m}$ integration, we follow Figure \ref{fig:domain} by splitting the domain for $\Sigma > 2$ and $\Sigma < 1$, corresponding to where the variables $\abs{\mathbf n}$ and $\abs{\mathbf m -\mathbf n}$ are large and small.
We also retain the $h_{nm}$ contribution instead of $g_{nm}$ for small $\abs{\mathbf m}$.
The contribution to the frequency shift in the small scale region $\Sigma > 2$ is given by
\begin{align}
    &-\frac{i\pi}{2\Delta} \int dn_x dn_y \, h_{nm}g_{mn} J_{nm}^2 b_{m-n}^2 \, .
\end{align} 
We notice on substitution, to this order in our asymptotics, that the only differences between this calculation and the large $\abs{\mathbf m}$ case are the absent factors of $-\abs{\mathbf m}$ from $g_{nm}$ and the prefactors $\abs{\mathbf m}^2$.
So the contribution to the frequency shift from small scales for small $\abs{\mathbf m}$ is
\begin{align}
    -\frac{\pi^2i }{256\Delta}\abs{\mathbf m }^2 \int_{2}^\infty \Sigma^2 b_{m-n}^2 d\Sigma \, .
\end{align}

In the large scale $\Sigma < 1$ regime, we find additional factors of $3/2$ in $g_{mn}$ and $\Sigma/2$ in $h_{nm}$.
As a result, we write down the contribution to the frequency shift as 
\begin{align}
    -\frac{3\pi^2 i}{1024\Delta} \abs{\mathbf m}^2 \int_{\abs{\mathbf m}}^2 \Sigma^3 b_{m-n}^2 \, d\Sigma \, .
\end{align}

For reference, the total perturbation energy given by integration over sum and difference coordinates is 
\begin{align}
    E &= \int_{\abs{\mathbf m}}^\infty d\Sigma \int_{-1}^1 dr \, \frac{\Sigma }{2\sqrt{1-r^2}} b_{m-n} \\ \nonumber
    &= \frac{\pi}{2}\int_{\abs{\mathbf m}}^\infty \Sigma b_{m-n}^2 \, d\Sigma \, .
\end{align}

\begin{acknowledgments}
\noindent This work was supported by U.S. Department of Energy Grant No. DE-FG02-04ER-54742.
\end{acknowledgments}

\nocite{*}

\bibliographystyle{unsrt}
\bibliography{hmhdnldamping}

\end{document}